# Current-Induced Magnetic Domain-Wall Motion by Spin Transfer Torque: Collective Coordinate Approach with Domain-Wall Width Variation

**Soon-Wook Jung and Hyun-Woo Lee**

Department of Physics, Pohang University of Science and Technology, Pohang, Kyungbuk 790-784, Korea

**The spin transfer torque generated by a spin-polarized current can induce the shift of the magnetic domain-wall position. In this work, we study theoretically the current-induced domain-wall motion by using the collective coordinate approach [Gen Tatara and Hiroshi Kohno, Phys. Rev. Lett. 92, 86601 (2004)]. The approach is extended to include not only the domain-wall position and the polarization angle changes but also the domain-wall width variation. It is demonstrated that the width variation affects the critical current.**

## 1. Introduction

Magnetization dynamics driven by a spin-polarized electric current is a fascinating subject and has attracted much interest since it was first predicted by Berger [1] and Slonczewski [2]. Notable examples are the magnetization reversal and the spin wave excitation in a nanomagent by a spin-polarized current. The origin of the current induced magnetization dynamics is the exchange interaction between conduction electron spins and the local magnetization. The spin angular momentum carried by conduction electrons is absorbed to the local magnetization and generates the spin transfer torque acting on the magnetization through the s-d exchange interaction. This current-induced spin torque can affect magnetic order, induce magnetic excitations, and even drive domain-wall (DW) motion. The idea of the current-induced domain-wall motion (CIDWM) studied by Berger [3] was later confirmed by several experiments [4] and followed by more theoretical works [5-7].

Li and Zhang [5] studied the CIDWM by using the LLG equation which includes the spin transfer torque term derived from their phenomenological theory assuming the adiabatic process. An alternative theoretical approach to the CIDWM was explored by Tatara and Kohno (TK) [6], who introduced the collective coordinates such as the wall center position $X$ and polarization $\phi_0$ (the angle between spins at the wall center and easy plane) and examined the CIDWM in terms of the collective coordinates. They derived the equations of motion for $X$ and $\phi_0$ from an effective Lagrangian for the system and analyzed the CIDWM quantitatively. However, there are indications that the DW motion is usually accompanied with the DW deformation [5,7] and thus it is not sufficient to describe the CIDWM using the variable $X$ and $\phi_0$ only. In this paper, we extend the collective coordinate approach, so that not only the variation of $X$ and $\phi_0$ but also the variation of the DW width $\lambda$ is taken into account. Use of the three collective coordinates ($X$, $\phi_0$, $\lambda$) makes the description for the DW motion more detailed.

## 2 . Collective coordinate approach

We consider a ferromagnetic wire consisting of a localized spin $\mathbf{S}$ with magnitude $S$ at each lattice site. It is assumed that the magnet have an easy-axis anisotropy along the z-axis with an anisotropy constant $K > 0$ and a hard-axis anisotropy constant $K_p > 0$. In the continuum limit, the Lagrangian for the localized spins $\mathbf{S} = S(\sin\theta\cos\phi, \sin\theta\sin\phi, \cos\theta)$ is given by [8]

$$L_S = \int \frac{d^3x}{a^3}\left[\hbar S \frac{d\phi}{dt}(\cos\theta - 1)\right] - H_S, \qquad (1)$$

$$H_S = \int \frac{d^3x}{a^3} \frac{S^2}{2} \left[ J \left\{ \left(\frac{\partial \theta}{\partial x}\right)^2 + \sin^2\theta \left(\frac{\partial \phi}{\partial x}\right)^2 \right\} + \sin^2\theta \left(K + K_p \sin^2\phi\right) \right], \quad (2)$$

where $a$ is the lattice constant, $H_S$ is the Hamiltonian of the spins and $J$ represents the Heisenberg exchange interaction between the localized spins. Here we have assumed that the spins are uniform in the $yz$ plane perpendicular to the $x$-axis along the wire, and we thus consider effective 1D model. In the absence of a current, we consider a DW configuration of the static solution which is related to minima of the Lagrangian $L_S$. As a DW profile, we take the standard Néel wall ($\phi = 0$); $\theta(x) = \cos^{-1} \tanh(x/\lambda_0)$, where $\lambda_0 = \sqrt{J/K}$ is the DW width.

The localized spins are coupled to conduction electrons via the s-d exchange interaction

$$H_{sd} = -\frac{\Delta}{S} \int d^3x \mathbf{S}(x) \left(c^\dagger \boldsymbol{\sigma} c\right), \quad (3)$$

where $2\Delta$ is the energy splitting due to the s-d exchange interaction. Here $c^\dagger (c)$ is the creation (annihilation) operator of conduction electrons subjected to the Hamiltonian $H_{el} = \sum_{\mathbf{k},\sigma} \varepsilon_{\mathbf{k}} c^\dagger_{\mathbf{k},\sigma} c_{\mathbf{k},\sigma} + H_{sd}$ with $\varepsilon_{\mathbf{k}} = \hbar^2 k^2 / 2m$. The spin-polarization of conduction electrons can change its direction through the s-d exchange interaction. This change leads to a torque (conventionally called spin transfer torque) acting on the localized spins, which induces the DW dynamics. In order to treat the motion of the Néel wall under the spin transfer torque, we consider the effective Lagrangian $L_{eff} = L_S - H_{ST}$, where $H_{ST} = -(\Delta/S) \int d^3x \mathbf{S}(x) \langle c^\dagger \boldsymbol{\sigma} c \rangle$ is the average of $H_{sd}$ with respect to conduction electrons.

In the collective coordinate approach [6], it is assumed that the essential part of the $\mathbf{S}(x,t)$ dynamics can be described by a small number of collective coordinates. In this work, we assume that $\theta(x,t)$ and $\phi(x,t)$ are related to three collective coordinates $\phi_0(t)$, $X(t)$, and $\lambda(t)$ via the following relations,

$$\phi = \phi_0(t), \qquad \theta(x,t) = \cos^{-1} \tanh\left(\frac{x - X(t)}{\lambda(t)}\right). \quad (4)$$

Physically speaking, Eq. (4) amounts the assumption that during the DW dynamics, the DW remains in the Néel wall form except for the possible changes of the DW position (parameterized by $X(t)$), the polarization angle (parameterized by $\phi_0(t)$), and the DW width (parameterized by $\lambda(t)$). Substituting

Eq. (4) into the effective Lagrangian $L_{eff}$, we obtain the Lagrangian for $\phi_0(t)$, $X(t)$, and $\lambda(t)$,

$$L_{eff} = -\frac{S^2 A}{a^3}\left[\frac{2\hbar}{S} X \frac{d\phi_0}{dt} + \frac{J}{\lambda} + \left(K + K_p \sin^2\phi_0\right)\lambda\right] - H_{ST}, \tag{5}$$

where $A$ is the cross sectional area of the wire.

The equations of motion for the three dynamic variables of the Néel wall in the presence of $H_{ST}$ can be derived from the Euler-Lagrange equation. For example, for $X(t)$,

$$\frac{d}{dt}\frac{\partial L_{eff}}{\partial \dot{X}} - \frac{\partial L_{eff}}{\partial X} = -\frac{\partial R}{\partial \dot{X}}, \tag{6}$$

where

$$R = \alpha \frac{\hbar}{2S}\int \frac{d^3 x}{a^3}\left(\frac{dS}{dt}\right)^2, \tag{7}$$

is a Rayleigh dissipation functional with the Gilbert damping parameter $\alpha$. It follows that we obtain the following equations of motion,

$$\frac{2\hbar S A}{a^3}\left(\frac{d\phi_0}{dt} + \frac{\alpha}{\lambda}\frac{dX}{dt}\right) = F_{el}, \tag{8}$$

$$\frac{2\hbar S A}{a^3}\left(\frac{dX}{dt} - \alpha\lambda \frac{d\phi_0}{dt}\right) = \frac{S^2 A K_p}{a^3}\lambda \sin 2\phi_0 + T_{el}, \tag{9}$$

$$\frac{S^2 A J}{a^3 \lambda^2} - \frac{\pi^2 \alpha \hbar S A}{6 a^3 \lambda}\frac{d\lambda}{dt} = \frac{S^2 A}{a^3}\left(K + K_p \sin^2 \phi_0\right) + F_\lambda, \tag{10}$$

where

$$F_{el} = \frac{\Delta}{S}\int d^3 x \frac{\partial S_0}{\partial X}\cdot \mathbf{n}(x), \quad T_{el} = -\frac{\Delta}{S}\int d^3 x \frac{\partial S_0}{\partial \phi_0}\cdot \mathbf{n}(x), \quad F_\lambda = -\frac{\Delta}{S}\int d^3 x \frac{\partial S_0}{\partial \lambda}\cdot \mathbf{n}(x). \tag{11}$$

Here $S_0$ represents $S$ with $\phi = \phi_0$, $\theta = \theta_0(x - X)$ and $\mathbf{n}(x) = \langle c^\dagger \boldsymbol{\sigma} c\rangle$ is the spin-density of the conduction electrons.

To complete the equations of motions (8), (9), and (10), $F_{el}$, $T_{el}$, and $F_\lambda$ need to be evaluated, which in turn requires the calculation of $\mathbf{n}(x)$. To calculate $\mathbf{n}(x)$, it is convenient to perform a local gauge transformation, $c(x) = U(x) a(x)$, where $U(x) = \mathbf{m}(x)\cdot \boldsymbol{\sigma}$ is an SU(2) matrix with $\mathbf{m} = \left(\sin(\theta_0/2)\cos\phi_0, \sin(\theta_0/2)\sin\phi_0, \cos(\theta_0/2)\right)$ and $a(x)$ is the annihilation operator in the rotated frame in the spin space. Then the spin-density $\mathbf{n}(x)$ can be expressed in terms of $\tilde{\mathbf{n}}(x) \equiv \langle a^\dagger \boldsymbol{\sigma} a \rangle = -i\text{Tr}\left(G^<(xt,xt)\boldsymbol{\sigma}\right)$, where $G^<_{\sigma,\sigma}(xt,x't') = i\langle a^\dagger_{x',\sigma'}(t')a_{x,\sigma}(t)\rangle$ is the lesser component of the Keldysh-Green function;

$$n_x(x) = [(1-\cos\theta_0)\cos^2\phi_0 - 1]\tilde{n}_x + (1-\cos\theta_0)\cos\phi_0\sin\phi_0\tilde{n}_y + \sin\theta_0\cos\phi_0\tilde{n}_z, \qquad (12a)$$

$$n_y(x) = (1-\cos\theta_0)\cos\phi_0\sin\phi_0\tilde{n}_x + [(1-\cos\theta_0)\sin^2\phi_0 - 1]\tilde{n}_y + \sin\theta_0\sin\phi_0\tilde{n}_z, \qquad (12b)$$

$$n_z(x) = \sin\theta_0\cos\phi_0\tilde{n}_x + \sin\theta_0\sin\phi_0\tilde{n}_y + \cos\theta_0\tilde{n}_z. \qquad (12c)$$

All three quantities $F_{\text{el}}$, $T_{\text{el}}$, and $F_\lambda$ can be expressed in terms of $\tilde{\mathbf{n}}(x)$. For instance, $F_\lambda$ becomes

$$F_\lambda = \Delta\int d^3x \frac{\partial\theta_0}{\partial\lambda}\left(\cos\phi_0\tilde{n}_x + \sin\phi_0\tilde{n}_y\right), \qquad (13)$$

which, in the Fourier space, can be written as

$$F_\lambda = A\Delta\int dq w_q\left(\cos\phi_0\tilde{n}_x(-q) + \sin\phi_0\tilde{n}_y(-q)\right), \qquad (14)$$

where

$$\tilde{\mathbf{n}}(q) = -i\int d^3k\text{Tr}\left[G^<(k+q,k)\boldsymbol{\sigma}\right], \qquad (15)$$

$$w_q = \int dx e^{-iqx}\frac{\partial\theta_0}{\partial\lambda} = -i\frac{\pi^2}{2}e^{-iqX}\frac{\sinh(\pi q\lambda/2)}{\cosh^2(\pi q\lambda/2)}. \qquad (16)$$

To calculate the Greens function, we consider the Hamiltonian of the electron part $H_{\text{el}}$ in the rotated frame which can be written as

$$H_{\text{el}} = \sum_{\mathbf{k},\sigma}\varepsilon_{\mathbf{k}\sigma}a^\dagger_{\mathbf{k}\sigma}a_{\mathbf{k}\sigma} + \frac{\hbar^2}{2m}\sum_{\substack{\mathbf{k},q \\ \sigma,\sigma'}}(2k+q)_x a^\dagger_{\mathbf{k}+q,\sigma}\left(\boldsymbol{B}(q)\cdot\boldsymbol{\sigma}\right)_{\sigma,\sigma'}a_{\mathbf{k},\sigma'}, \qquad (17)$$

where $\varepsilon_{\mathbf{k}\sigma} = \hbar^2 k^2/2m - \sigma\Delta$, $\boldsymbol{B}(x) = \int dq e^{iqx}\boldsymbol{B}(q)$ with $\boldsymbol{B}(x) = \mathbf{m}\times(\partial\mathbf{m}/\partial x)$. In Eq. (17), we have ignored the higher order terms of $\boldsymbol{B}(q)$ since we are interested in the thick DW ($\partial\mathbf{m}/\partial x$ is small). In this case, we obtain the Dyson equation in the Keldysh space, to the lowest order of $\boldsymbol{B}(q)$,

$$G(\mathbf{k}t,\mathbf{k}'t') = g(\mathbf{k},t-t')\delta_{\mathbf{k},\mathbf{k}'} + \int_C dt_1 g(\mathbf{k},t-t_1)\left[\frac{\hbar^2}{2m}(k+k')_x\left(\boldsymbol{B}(k-k')\cdot\boldsymbol{\sigma}\right)\right]g(\mathbf{k}',t_1-t'), \qquad (18)$$

where, $C$ is denoted by contour integral and $g$ is the free Green function [9]. Then the lesser component

of the Keldysh-Green functions is given by [9]

$$G^<(kt,\mathbf{k}'t') = g^<(\mathbf{k},t-t')\delta_{\mathbf{k},\mathbf{k}'} + \frac{\hbar^2(k+k')_x}{2m}\int_{-\infty}^{\infty}dt_1\left[g^r(\mathbf{k},t-t_1)(\mathbf{B}\cdot\boldsymbol{\sigma})g^<(\mathbf{k}',t_1-t') + g^<(\mathbf{k},t-t_1)(\mathbf{B}\cdot\boldsymbol{\sigma})g^a(\mathbf{k}',t_1-t')\right]. \quad (19)$$

In the Fourier space, we can obtain

$$G^<(\mathbf{k},\mathbf{k}';\omega) = g^<(\mathbf{k},\omega)\delta_{\mathbf{k},\mathbf{k}'} + \frac{\hbar^2(k+k')_x}{2m}\left[g^r(\mathbf{k},\omega)\left(\mathbf{B}(\mathbf{k}-\mathbf{k}')\cdot\boldsymbol{\sigma}\right)g^<(\mathbf{k}',\omega) + g^<(\mathbf{k},\omega)\left(\mathbf{B}(\mathbf{k}-\mathbf{k}')\cdot\boldsymbol{\sigma}\right)g^a(\mathbf{k}',\omega)\right], \quad (20)$$

where the unperturbed Green functions are given by $g^<_{\sigma,\sigma'}(\mathbf{k},\omega) = 2\pi i f(\omega)\delta(\omega-\omega_{\mathbf{k},\sigma})\delta_{\sigma,\sigma'}$ and $g^{r,a}_{\sigma,\sigma'}(\mathbf{k},\omega) = \delta_{\sigma,\sigma'}(\omega-\omega_{\mathbf{k},\sigma}\pm i0)^{-1}$ with $\varepsilon_{\mathbf{k},\sigma} = \hbar\omega_{\mathbf{k},\sigma}$. We get thus the electron spin-density,

$$\tilde{n}_x(-q) = \frac{i\hbar^2}{4\pi}\int\frac{d^3k}{(2\pi)^3}(2k+q)_x u_{-q}\sum_\sigma \sigma e^{-i\sigma\phi_0}\frac{f(\varepsilon_{\mathbf{k}+q,-\sigma})-f(\varepsilon_{\mathbf{k},\sigma})}{\varepsilon_{\mathbf{k}+q,-\sigma}-\varepsilon_{\mathbf{k},\sigma}+i0}, \quad (21a)$$

$$\tilde{n}_y(-q) = -\frac{\hbar^2}{4\pi}\int\frac{d^3k}{(2\pi)^3}(2k+q)_x u_{-q}\sum_\sigma e^{-i\sigma\phi_0}\frac{f(\varepsilon_{\mathbf{k}+q,-\sigma})-f(\varepsilon_{\mathbf{k},\sigma})}{\varepsilon_{\mathbf{k}+q,-\sigma}-\varepsilon_{\mathbf{k},\sigma}+i0}, \quad (21b)$$

where $u_q = -\int dx e^{-iqx}\frac{\partial\theta_0}{\partial x} = \frac{\pi e^{-iqX}}{\cosh(\pi q\lambda/2)}$. From the Eq. (14), $F_\lambda$ then becomes

$$F_\lambda = -\frac{iA\hbar^2\Delta}{2\pi}\int dq\int\frac{d^3k}{(2\pi)^3}\sum_\sigma w_q u_{-q}\sigma f_{\mathbf{k},\sigma}\frac{(2k+q)_x}{2m}\frac{P}{\varepsilon_{\mathbf{k}+q,-\sigma}-\varepsilon_{\mathbf{k},\sigma}}. \quad (22)$$

In a similar way, we obtain

$$F_{el} = \frac{A\hbar^2\Delta}{2}\int dq\int\frac{d^3k}{(2\pi)^3}\sum_\sigma u_q u_{-q}\sigma f_{\mathbf{k},\sigma}\frac{(2k+q)_x}{2m}\delta\left(\varepsilon_{\mathbf{k}+q,-\sigma}-\varepsilon_{\mathbf{k},\sigma}\right), \quad (23)$$

$$T_{el} = \frac{A\Delta\lambda}{2\pi}\int dq\int\frac{d^3k}{(2\pi)^3}\sum u_q u_{-q} f_{\mathbf{k},\sigma}\frac{(2k+q)_x}{2m}\frac{P}{\varepsilon_{\mathbf{k}+q,-\sigma}-\varepsilon_{\mathbf{k},\sigma}}. \quad (24)$$

In this paper, we are interested in the adiabatic limit where the DW width $\lambda(t)$ is much larger than the Fermi wavelength $k_F^{-1}$. In this case, $u_q u_{-q} \to (4\pi/\lambda)\delta(q)$, $(\varepsilon_{\mathbf{k}+q,-\sigma}-\varepsilon_{\mathbf{k},\sigma})_{q=0} = 2\sigma\Delta$, and it can be verified that

$$F_{el} = 0, \quad (25)$$

$$T_{el} = \frac{\hbar^2}{L}\sum_\mathbf{k}\sigma\frac{k_x}{m}f_{\mathbf{k},\sigma} = Aj_s, \quad (26)$$

where $j_s \equiv \frac{\hbar}{AL}\sum_{\mathbf{k}\sigma}\sigma\frac{\hbar k_x}{m}f_{\mathbf{k}\sigma}$ is conventionally defined as spin current density. Therefore, we can see that $T_{el}$ is the rate at which the spin is transported through a given area $A$. It is possible to control $T_{el}$ electrically in experiments since the electric current density $j_e$ is related to spin current density $j_s = (\hbar/e)pj_e$, where $p$ is a material parameter representing the spin-polarization of the current. Noting that

$$w_q u_{-q} = -\frac{1}{\pi\lambda}\frac{d}{dq}u_q u_{-q} \to \frac{2}{\lambda^2}\frac{d}{dq}\delta(q), \qquad \text{(for adiabatic limit)} \qquad (27)$$

$F_\lambda$ can be calculated as

$$F_\lambda = -\frac{A\hbar^2}{4m\lambda^2}\int\frac{d^3k}{(2\pi)^3}\sum_\sigma f_{\mathbf{k},\sigma} + \frac{A\hbar^2}{4\Delta\lambda^2}\int\frac{d^3k}{(2\pi)^3}\sum_\sigma \sigma\left(\frac{\hbar k_x}{m}\right)^2 f_{\mathbf{k},\sigma}$$

$$= -\frac{3}{20}\frac{A\hbar^2 n}{m\lambda^2}\left(1 - \frac{2(\eta^3-1)(\eta^2+1)}{3(\eta^3+1)(\eta^2-1)}\right), \qquad (28)$$

where $n$ is the electron density, and $\eta = k_{F\uparrow}/k_{F\downarrow} \geq 1$ is the spin polarization.

Finally we obtain the equations of motion for $X$, $\phi_0$, and $\lambda$ in the adiabatic DW limit,

$$(1+\alpha^2)\frac{d\phi_0(t)}{dt} = -\frac{\alpha}{\lambda(t)}\left[\frac{SK_p}{2\hbar}\lambda(t)\sin 2\phi_0(t) + v_{el}\right], \qquad (29)$$

$$-\frac{\pi^2\alpha\hbar}{6S\lambda(t)}\frac{d\lambda(t)}{dt} = -\frac{1}{\lambda(t)^2}(J+J_\lambda) + K + K_p\sin^2\phi_0(t), \qquad (30)$$

$$\frac{dX(t)}{dt} = \frac{1}{1+\alpha^2}\left[\frac{SK_p}{2\hbar}\lambda(t)\sin 2\phi_0(t) + v_{el}\right], \qquad (31)$$

where $J_\lambda \equiv -(a^3\lambda^2/S^2A)F_\lambda$ and $v_{el} \equiv (a^3/2\hbar S)j_s$. It is noted that $v_{el}$ has a dimension of velocity and represents the rate of spin transfer.

## 3. Results and Discussion

The Eqs. (29), (30), and (31) for $X$, $\phi_0$, and $\lambda$ describe the DW dynamics in the presence of an electric current. The numerical results for the equations are given in Fig. 1 and Fig. 2. The following parameters are used in the plot; $J = 3\times 10^{-40}\,\text{Jm}^2$, $K = 6.2\times 10^{-24}\,\text{J}$, $K_p = 3\times 10^{-23}\,\text{J}$, $\eta = 11$ [10], and $\alpha = 0.02$ for Co.

Fig. 1 shows the time averaged velocity of the DW center, $v_{av} \equiv \langle dX/dt\rangle$ as a function of the spin transfer $v_{el}$ proportional to the spin current. Note the presence of a threshold $v_{el}^c$. The DW is not driven for $v_{el} < v_{el}^c \approx 1000 \text{m/s}$. When $v_{el}$ is larger than this threshold, the average velocity $v_{av}$ of DW increases rapidly and for sufficiently larger $v_{el}$, $v_{av}$ becomes proportional to the spin transfer $v_{el}$. The motion of the DW is closely coupled with the DW deformation as shown in the Fig. 2. Before the current bias is applied, $v_{el} = 0$ and $\phi_0 = 0$. Immediately after the finite current bias is turned on, the spin transfer $v_{el}$ leads to the DW motion with initial velocity $v_0 = v_{el}/(1+\alpha^2)$. Simultaneously, the spin angular momentum from conduction electrons is transferred to the localized spins so that $\phi_0$ and $\lambda$ increase in time (DW deformation). For $v_{el} < v_{el}^c$, with $\phi_0$ and $\lambda$ saturated within a nanosecond, the spin angular momentum is completely absorbed to deform the DW and thus the wall motion stops eventually. When the spin transfer rate $v_{el}$ is lager than the absorption rate to $\phi_0$ and $\lambda$ for $v_{el} > v_{el}^c$, the net spin transfer with nonzero drives the stream motion and the DW does not stop. Note that this threshold behavior is similar to the transition (so called Walker breakdown) of the field-driven DW motion case, where the stationary canted angle $\phi_0$ below the threshold field can no more remain stationary, $d\phi_0/dt \neq 0$, leading to oscillatory velocity of the DW above the threshold [11]. However, while the field-driven DW motion is possible even below the threshold because it is driven by demagnetizing field torque due to the deformation ($\phi_0 \neq 0$), CIDWM is not possible since the spin transfer $v_{el}$ balances the DW deformation. Therefore, one can expect there is the intrinsic pinning potential related to the hard axis anisotropy $K_p$ in the CIDWM ( see Eq. (33) ).

This threshold behavior can be extracted analytically as well. For sufficiently small $v_{el}$, it is possible to find special $\phi_0$ and $\lambda$ that make the R.H.S. of both Eq. (29) and (30) zero. Note that for such $\phi_0$ and $\lambda$, the R.H.S. of Eq. (31) vanishes as well, describing the non-moving steady state of the DW for $v_{el} < v_{el}^c$. On the other hand, if the spin transfer $v_{el}$ is so large that the R.H.S. of Eq. (29) never become zero despite of the changes of $\phi_0$ and $\lambda$, the DW keeps on their stream motion. To find this threshold $v_{el}^c$, we consider the situation when $d\lambda/dt = 0$ and express $\lambda(t)$ in terms of $\phi_0(t)$. Then we obtain

$$\frac{SK_p}{2\hbar}\lambda(t)\sin 2\phi_0(t) = \frac{SK_p\sqrt{J+J_\lambda}}{2\hbar}\frac{\sin 2\phi_0(t)}{\sqrt{K+K_p\sin^2\phi_0(t)}}. \quad (32)$$

Note that once $v_{el}$ is lager than the maximum value of the R.H.S. of the Eq. (32), the R.H.S. of Eq. (31) can not vanish. Maximizing $v_{el}$ with respect to the angle $\phi_0$, we find

$$v_{el}^c = \frac{S}{\hbar}\sqrt{(J+J_\lambda)\left(2K+K_p-2\sqrt{K(K+K_p)}\right)}. \quad (33)$$

Its estimated value is about 1000m/sec corresponding to the electric current density $j_e = 2.6\times 10^{13}\,\mathrm{A/m^2}$ for the complete spin-polarization $p=1$. From the Eq. (33), we see that the critical value $v_{el}^c$ for CIDWM depends on the sample geometry because it is related to shape anisotropy, i.e. the hard axis anisotropy $K_p$. Since $K_p$ usually increase with decreasing the aspect ratio ($R=t/W$ where $t$ is the thickness and $W$ is the width) of a nanowire, one can control $v_{el}^c$ by changing sample geometry [12]. This effect is shown in the Fig. 3 as a function of the aspect ratio [13] setting $J_\lambda = 0$ for direct comparison with TK's result. The critical value $v_{el}^c$ is different from the TK's result [6] $v_{el}^c = SK_p\lambda_0/2\hbar$, which ignores the DW width variation. From the Eq. (32), we can see that the DW width variation can reduce the critical value $v_{el}^c$, corresponding to the maximum value of the Eq. (32). The effect of the DW width variation on $v_{el}^c$ becomes larger as $K_p$ increase, so that $v_{el}^c$ becomes smaller than the TK's result as $R$ decrease.

We remark that a similar behavior is obtained in Ref. [5]. Indeed, Eqs. (29), (30), and (31) are similar to the equations of motion derived in Ref. [5] from LLG equation with adiabatic spin torque term. However, Eq. (30) has quite different form; it includes "$d\lambda/dt$" term and "$J_\lambda$" term. The "$d\lambda/dt$" term arises from the energy relaxation due to the DW energy variation (Gilbert damping effect). In Ref. [5], it was ignored. The "$J_\lambda$" term arises from the quantum mechanical effect due to the s-d exchange interaction between the localized spins and the conduction electron spins, which was not considered in Ref. [5]. As seen in Eq. (30), the s-d exchange interaction leads to modify the Heisenberg exchange interaction $J$, from which we find that the initial DW width $\lambda_0 = \sqrt{J/K}$ should be increased to $\lambda = \sqrt{(J+J_\lambda)/K}$ due to the conduction electron spin. For the value $\eta = 11$, we estimate $J_\lambda \approx 5.9 \times 10^{-40}\, \text{Jm}^2$ which is of the same order as $J \approx 3 \times 10^{-40}\, \text{Jm}^2$.

Finally, we comment briefly on effects of the so-called nonadiabatic spin transfer torque [14] ignored in this paper. This additional spin transfer torque arises from the fact that the direction of the spin-polarization of the current does not follow the localized spin direction due to the spin flip scattering or contribution of the higher order perturbations. In Refs. [7] and [11], it is demonstrated that this additional spin transfer torque induces a small non zero DW speed for $v_{el} < v_{el}^c$. According to Ref. [11], however, the magnitude of the nonadiabatic spin torque is much smaller than that of the conventional spin transfer torque taken into account in this work and it is demonstrated in Ref. [7] that the relation between $v_{av}$ and $v_{el}$ (Fig. 1) is only minor affected by the nonadiabatic spin transfer torque.

In conclusion, we have studied the CIDWM in the adiabatic limit using the collective coordinate approach. Taking the variation of the DW width $\lambda$ as well as center position $X$ and polarization $\phi_0$, we have obtained the equations of motion for the DW motion and shown the relation between the DW motion and deformation. Moreover, we find that the change of the DW width can modify the critical current density for the wall motion.

## Acknowledgement

The authors thank Kyung-Jin Lee and Chun-Yeol You for valuable comments.

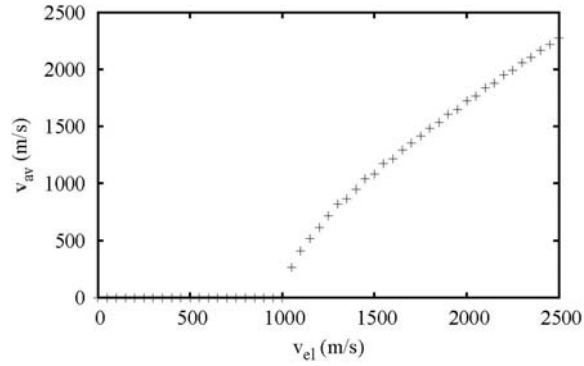

Fig. 1. Time averaged wall velocity $v_{av}$ as function of spin transfer $v_{el}$.

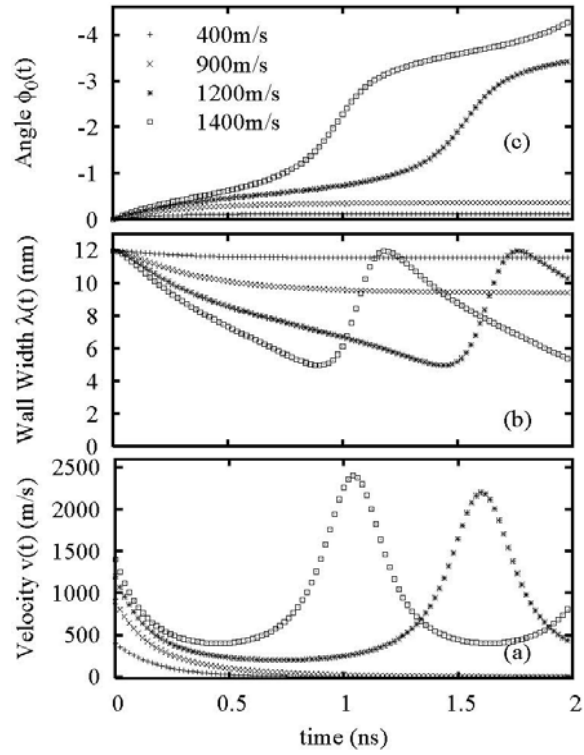

Fig. 2. The DW polarization $\phi_0$, the wall width $\lambda$, and the velocity $v$ as function of time for various spin current $v_{el}$=400m/s, 900m/s, 1200m/s, and 1400m/s. For $v_{el} < v_{el}^c \approx 1000\text{m/s}$, $\phi_0$ and $\lambda$ are saturated at about one nanosecond, and the DW motion stops. For $v_{el} > v_{el}^c$, the DW moves with the velocity oscillating around the initial velocity.

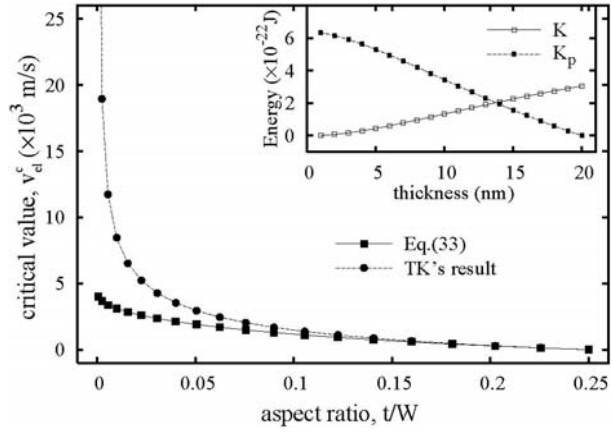

Fig. 3. The critical value $v_{el}^c$ as function of aspect ratio (=$t/W$ where $t$ is thickness and $W$ is width) of nanowire with constant cross-sectional area, $A$=1600nm$^2$ where the transverse wall is stable. The inset shows the relation between the easy axis anisotropy $K$, the hard axis anisotropy $K_p$ and the thickness $t$.